\documentclass[12pt,preprint]{aastex}
\usepackage{natbib}
\shorttitle{Scaling Law Model for the Planetary Magnetism}
\shortauthors{F. X. Alvarez}

\begin{document}

\title{Scaling Law for the Magnetic Field of the Planets
Based on a Thermodynamic Model}

\author{F. X. Alvarez}
 \email{Xavier.Alvarez@uab.es}
\affil{Departament de F\'isica, Universitat Aut\`onoma de
Barcelona, 08193 Bellaterra, Catalonia, Spain}


\begin{abstract}
A thermodynamic model for the generation of magnetic fields in the
planets is proposed, considering crossed effects between
gravitational and electric forces. The magnetic field of the Earth
is estimated and found to be in agreement with the actual field.
The ratio between the field of several planets and that of the
Earth is calculated in the model and compared with the same ratio
for the measured fields. These comparisons are found to be
qualitatively consistent. Once the value of the magnetic field is
calculated, the model is used to obtain the tilt of the magnetic
dipole with respect to the rotation axis. This model can explain
why Uranus and Neptune magnetic fields have higher quadrupole
moment than the other magnetic fields of the Solar System and why
Saturn, that has a highly axysymmetric field, has lower
quadrupolar component. The model also explains the double peak of
the magnetic field observed by Voyager 2 while recording the field
of Neptune. The Earth paleomagnetic data are analysed and found to
be consistent with the model, that predicts higher quadrupole
components for the more tilted dipoles. A field is predicted for
all the planets and satellites of the Solar System with enough
mass. Objections are made to the theories that predict that this
effect could not generate a field agreeing with the measured one.
\end{abstract}

\keywords{MHD, planets and satellites: general, magnetic fields}
\maketitle

\include{ms.bib}

\section{Introduction}

One of the open fields in modern physics is to find the equations
that rule the origin and evolution of the magnetic field of the
planets and, specifically, that of the Earth. Some theories based
on the dynamo model have obtained a correct value for the magnetic
field for the Earth (\cite{GLATZ4,BUS}), and even some of the
proposed models describe a magnetic pole inversion as observed in
the Earth (\cite{GLATZ1, GLATZ3, PLUN}). The dynamo model has also
been applied to describe the magnetic field of the Sun
(\cite{GLATZ2}). Even though these theories seem to fit correctly
to the field of our planet, there is a question still unsolved.
Why all the objects of the Universe seem to be closely related
with magnetic fields?. Six of the nine planets of our solar
system, main sequence stars, neutron stars, galaxies, etc..., all
seem to have high magnetic fields.

If we look in further detail to the Solar System, the presence of
significative magnetic fields seems to be the law rather than the
exception (\cite{RUSSELL1,RUSSELL2}). But this ubiquity in the
presence of magnetic fields does not seem to remain if we look at
the morphologies of such fields. Mercury and Jupiter show
Earth-like structure fields (\cite{NESS1}). Uranus
(\cite{CONNEREY1}) and Neptune very inclined fields and
quadrupolar moments. In the case of Neptune, the Voyager 2 probe
recorded a double peak in the mesured magnetic field
(\cite{CONNEREY2}). Saturn, in contrast, has a highly axisymmetric
magnetic dipolar field (\cite{CONNEREY3}).

Although the geodynamo model explains so well the Earth magnetic
field, it must make nontrivial assumptions to incorporate subtle
details in the magnetic fields of other planets. For instance,
Uranus and Neptune's magnetic fields are more quadrupolar than the
one predicted by the outer core convective dynamo. In the same
manner paleomagnetic data in the Earth show an increment of the
relative intensity of the quadrupolar components of the magnetic
fields during the magnetic dipole reversals. Another problem is
that Saturn's strongly axisymmetric field contradicts Cowling's
theorem that no axisymmetric homogeneous magnetic field can be
self-sustained. There have been solutions for each of these
problems. The quadrupolar problem of Uranus and Neptune has been
solved by a change of the convection zone. In the modified model,
the convection is produced in a zone between \(0.75\) and \(0.8\)
times the radius of the planet (\cite{AUR, STAN}). The model
proposed uses a fluid electrically-conducting inner core, instead
of the solid electrically-conducting inner core of the standard
geodynamo models. The high axysimmetry of Saturn has been solved
in models like \cite{STEVENSONSAT}. All these solutions make the
theory not easily applicable to all different cases in the Solar
System with one simple model. Although dynamo theory is the best
accepted way to fit to the experimental data, the models used to
describe these facts are different for every object described.
This paper tries to solve this problem with the use of a model
that tries to explain all these features without the need
specificities.

In this paper we propose a model of magnetic field generation
based on a thermodynamic point of view which uses the relation
between thermodynamic forces and fluxes as the starting point. In
the model, a redistribution of the charge is obtained from the
gravitational energy of the planet through the action of pressure.
From this charge distribution and the rotation of the planet, the
magnetic field is generated. Several advantages of this model are:
1) Its ability to spontaneously break the spherical symmetry of
the system to yield a field with axial symmetry; 2) the
possibility to obtain a tilt of the dipole moment relative to the
axis of rotation; 3) a possible explanation of why Uranus and
Neptune, that have large tilts, have higher quadrupole moment than
the other planets; 4) an explanation of why Saturn, with a high
axisymmetric field, has a low quadrupole moment.

The possibility of reversing the dipole is open, and predicts that
the field during a transition may become more quadrupolar than the
stable normal or reverse state (as observed in the past). The
detailed dynamics of the transition is out of the scope of this
paper that only pretends to present the main lines of the model.

The inclusion of an analysis of the multipolarity of the fields
obtained is also out of the scope of this article, but this
feature is implicitily considered in the model because of its
geometry.

This paper doesn't attempt to replace the geodynamo model that
fits so well in the case of the Earth (\cite{GLATZ4, GLATZ5}), but
tries to propose a general mechanism for the different planets to
generate a magnetic field. Instead of it, the energy needed for
the generation of the field comes from the more universal
gravitational field, present all over the Universe. To describe in
an unified way the behaviors of such different planets, it is
logical to ask for a model relatively independent of the
microscopic details, this is the reason why we have turned here to
a global thermodynamic analysis, at a macroscopic level, rather
than directly going to the mechanistic details, which may be
different in different planets.

The model could be used in posterior papers, as a generator of
magnetic fields in more general objects of the Universe like the
main sequence stars or the neutron stars.

\section{Model}

We start our discussion by calculating the conditions of pressure
in the core. The pressure inside the planet is given in the first
approximation by the hydrostatic equilibrium between gravitation,
that pushes inwards, and pressure, that pushes outwards. The
resulting value of the pressure distribution is
\begin{eqnarray}
\label{hydrostatic}
p(r)=\frac{9GM^2}{8\pi}\Bigg(\frac{1}{R^4}-\frac{r^2}{R^6}\Bigg),
\end{eqnarray}
where \(M\) and \(R\) are the total mass and external radius of
the planet and \(r\) is the radius at the position being
considered. In this relation it is assumed that the density is
constant and equal to the average density of the planet. This
prevents the gravitational collapse of the planet. We must take
into account that we don't pretend to have an exact value for this
parameter, but only an idea of its magnitude and we don't need a
more accurate value of it.

The internal pressure of the Earth makes iron in the inner part of
the core to be more compact packed than iron in the outer part.
This increase in atomic packing makes the Fermi energy in the
inner part of the core to be higher than the one in the outer
part. This difference in the Fermi energy pushes the electrons in
the conduction band outside the center of the Earth. If the charge
of the materials were zero (conducting iron in the inner core and
electrons pushed outside), this tendency would continue until the
complete separation of both particles, but the particles are
charged and we face a problem of stratification with electrical
charge. In this scenario, there is no need of ionization of the
core's material, the only assumption made is its conductivity, and
the fact that the conduction electrons have complete mobility in
all the conductor. The property of conductivity is common to all
the planet because in the core all the planets have high enough
pressures to cause the constituent material to act as a metal even
though the material is hydrogen like in Jupiter and Saturn.

Thermodynamically we could see the model as a gradient of chemical
potential due to a pressure gradient. Electrons in a material are
subject to the force exert by the gradient of the electrochemical
potential. This gradient has the form
\begin{eqnarray}
\nabla \mu_{el}= \mathbf{E}+\nabla \mu,
\end{eqnarray}
where \(\mathbf{E}\) is the external electric field and
\(\mu_{el}\) and \(\mu\) are respectively the electrochemical and
chemical potentials.

A gradient of chemical potential could be caused by several
reasons. We could obtain a gradient by changing the concentration,
the pressure and the temperature. Thermodynamically, the cause of
an electric current is the electrochemical gradient. If there is
no current present in a material, the gradient of electrochemical
potential must be zero. If we take into account all these
generators in the last relation and impose the equilibrium
condition, we obtain
\begin{eqnarray}
\frac{\partial \mu}{\partial p} \nabla p+\frac{\partial
\mu}{\partial T} \nabla T+\frac{\partial \mu}{\partial c} \nabla
c=0,
\end{eqnarray}
where \(p\) is the pressure, \(T\) temperature and \(c\)
concentration.

In the radial direction in the Earth core, the conservation of
charge grants that the net current over sufficient long periods of
time must be zero. This condition leave only the possibility that
the radial current is zero or oscillate around this value. We
start our discussion by taking the firs possibility. The main
cause of 

In this simple model we assume two concentric spherical shells
filled with some substance with electric permittivity \(\epsilon\)
, electric conductivity \(\sigma\) , and thermal conductivity
\(\kappa\) . These shells corresponds to the inner core boundary
(ICB) and the core mantle boundary (CMB).

The energy balance equation of the material filling the system is,
if we consider electrical effects,
\begin{eqnarray}
\label{ener} \rho \dot{u}=- \nabla \cdot \mathbf{q} + \mathbf{i}
\cdot \mathbf{E},
\end{eqnarray}
where \(\rho\) is the mass density, \(u\) internal energy per unit
mass, \(\mathbf{i}\) electrical current density, \(\mathbf{q}\)
the energy flux and \(\mathbf{E}\) the electric field. The
conservation of mass of component \(k\) is expressed by
\begin{eqnarray}
\label{consma} \rho \dot{c_k}=- \nabla \cdot \mathbf{j_k},
\end{eqnarray}
with \(c_k=\rho_k/\rho\) the mass fraction and \(\mathbf{j_k}\)
the material current density.

The classical entropy of the system is defined with the usual
Gibbs equation
\begin{eqnarray}\label{entro1}
ds = \frac{1}{T}du - \frac{p}{T} dv -
\sum_{k=1}^{n}\frac{\mu_k}{T}dc_k.
\end{eqnarray}
By differentiating this equation with respect to time, and
substituting (\ref{ener}) and (\ref{consma}) in the subsequent
equation we obtain
\begin{eqnarray}\label{entro2}
\rho\dot{s} & =  & \frac{1}{T}\Big(- \nabla \cdot \mathbf{q} +
\mathbf{i} \cdot \mathbf{E}\Big) - \frac{p\rho}{T} \dot{v} -
\sum_{k=1}^{n}\frac{\mu_k}{T}\nabla \cdot \mathbf{j_k}.
\end{eqnarray}
Arranging terms, we obtain four fluxes and four generalized forces
appearing in the entropy production, namely, the last four terms
in the right hand side of
\begin{eqnarray}\label{entro3b}
\rho\dot{s} & = & - \nabla \cdot \Big(\frac{\mathbf{q -
\sum_{k=1}^{n} \mu_k \mathbf{j_k}}}{T} \Big) + \Big(\mathbf{q -
\sum_{k=1}^{n} \mu_k \mathbf{j_k}}\Big)\cdot\nabla \Big(\frac{\mathbf{1}}{T}\Big)+ \\
\nonumber & & +\mathbf{i} \cdot \frac{\mathbf{E}}{T} - \frac{\rho
p}{T} \dot{v} - \sum_{k=1}^{n}\frac{\nabla\mu_k}{T}\cdot
\mathbf{j_k}
\end{eqnarray}
\begin{table}[h] \centering
\begin{tabular}{|c|c|}     \hline
\textbf{Flux} & \textbf{Generalized force} \\
\hline
\(\dot{v}\)           & \(\rho p/T\)                        \\
\(\mathbf{j_k}\)      & \(\nabla \mu_k/T\)                        \\
\(\mathbf{q -
\sum_{k=1}^{n} \mu_k \mathbf{j_k}}\)           & \(\nabla(1/T)\)                        \\
\(\mathbf{i}\)      & \(\mathbf{E}/T\)                        \\
\hline   \end{tabular}   \caption{table showing the fluxes and the
corresponding generalized forces appearing in the entropy
production in Eq (\ref{entro3b})}\label{tabflux}
\end{table}

As it is known, a generalized force can produce not only its
associated flux, but also other fluxes, like in thermoelectricity
or thermodiffusion (\cite{GROOT, EIT1}). In our example, we
analyse the coupling between electric field and the gradient of
chemical potential. The constitutive equations of the
corresponding fluxes are
\begin{eqnarray}
\mathbf{i} = L_{ee} \frac{\mathbf{E}}{T} + L_{eg} \frac{\nabla \mu_k}{T}, \\
\mathbf{j_k} = L_{ge} \frac{\mathbf{E}}{T} + L_{gg} \frac{\nabla
\mu_k}{T},
\end{eqnarray}
where \(L_{ee}\) is related to the electrical conductivity of the
material (\(L_{ee}=\sigma T\)), \(L_{gg}\) is related to the
diffusion coefficient of component \(k\) and \(L_{eg}=L_{ge}\) is
the crossed coefficient that relates the electric and
gravitational effects. We use the \(g\) subscripts because, below,
this value will be related with the gravitational energy.

The gradient of the chemical potential is related with the
pressure gradient as
\begin{eqnarray}
\nabla \mu_k = v_k \nabla p,
\end{eqnarray}
where \(v_k\) is the specific molar volume of species k. We
consider that the dependence of \(\mu_k\) on the temperature is
negligible as compared with that on the pressure. Using this
relation and an approximate value of the gradient of \(p\)
obtained from (\ref{hydrostatic}), we get the following relation
between the fluxes,
\begin{eqnarray}
\mathbf{i} = L_{ee} \frac{\vec{\mathbf{E}}}{T} + L_{eg} \frac{v_k}{T}\Bigg(\frac{9GM^2}{8\pi}\frac{1}{R^5}\Bigg)\mathbf{\hat{r}}, \\
\mathbf{j_k} = L_{ge} \frac{\vec{\mathbf{E}}}{T} + L_{gg}
\frac{v_k}{T}\Bigg(\frac{9GM^2}{8\pi}\frac{1}{R^5}\Bigg)\mathbf{\hat{r}},
\end{eqnarray}
with \(\mathbf{\hat{r}}\) the unit vector in the radial direction.

If we consider that the planet interior is in equilibrium, the
radial electrical and material currents must vanish, in such a way
that there must be an electrical field that opposes to the
gravitational effects on the charged fluid
\begin{eqnarray}
L_{ee} \frac{\vec{\mathbf{E}}}{T} + L_{eg}
\frac{v_k}{T}\Bigg(\frac{9GM^2}{8\pi}\frac{1}{R^5}\Bigg)\mathbf{\hat{r}}=0,
\end{eqnarray}
which yields
\begin{eqnarray}\label{magneticfield}
\vec{\mathbf{E}}=- \frac{L_{eg}}{L_{ee}}\nabla \mu= -
\frac{L_{eg}}{L_{ee}}
v_k\Bigg(\frac{9GM^2}{8\pi}\frac{1}{R^5}\Bigg)\mathbf{\hat{r}}.
\end{eqnarray}

\begin{figure}
\label{plotone}
\includegraphics{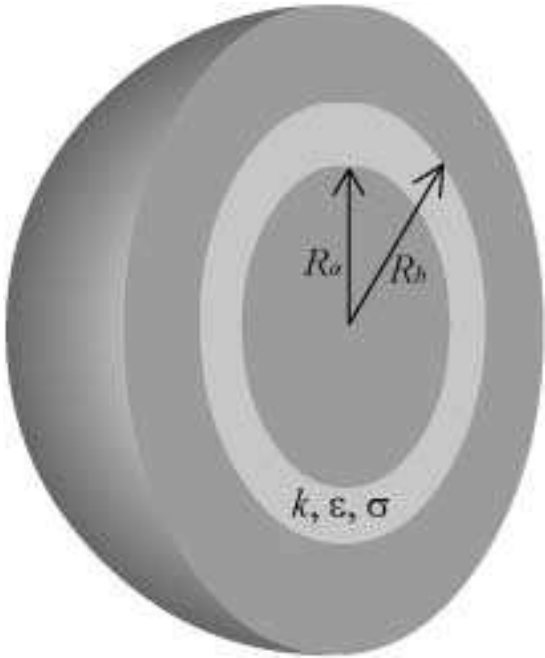}
\caption{Sketch of the interior of a planet with the symbols being
used in the text}
\end{figure}

\begin{figure}
\label{plottwo}
\includegraphics[scale=0.6]{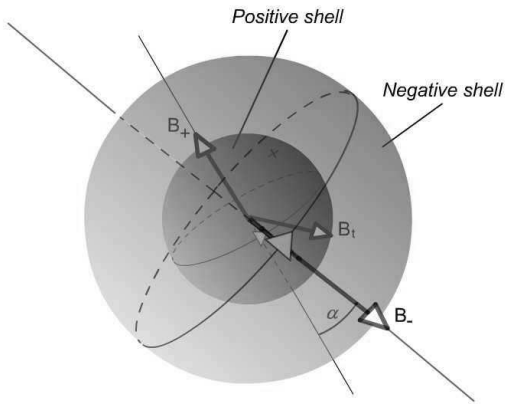}
\caption{Image of the magnetic field created by each one of the
different shells and the total magnetic field.}
\end{figure}

Now it is time to examine the effect of this electrical field on
the electrons of the core. Figure \ref{plotone} shows the model we
use. We can see it like a condenser composed of two spherical
plates of radius \(r_a=\alpha R\) (inner sphere) and \(r_b=\beta
R\) (outer sphere), where \(R\) is the planet radius and
\(\alpha\) and \(\beta\) are numerical coefficients which take a
value between \(0\) and \(1\). We suppose that the planet stores
electrical charge until the field created by this charge cancels
the one created by gravitational effects.

To calculate the charge we need the value of the capacity of the
condenser (\cite{WANGSNESS}). This value is

\begin{eqnarray}
\frac{1}{C} = \frac{1}{4\pi
\epsilon}\Big(\frac{1}{r_a}-\frac{1}{r_b}\Big)=\frac{1}{4\pi
\epsilon}\frac{1}{R}\Big(\frac{\beta - \alpha}{\beta \alpha}\Big)
\end{eqnarray}

The value of the electrical field created by this condenser is
related with the charge stored in it. If we suppose that the value
of the electric field is approximately constant inside the plates
and it is given by (\ref{magneticfield}) we obtain for the charge

\begin{eqnarray}\label{CARREGA}
Q & =  & C  \Delta V = C \mathbf{E} (r_b - r_a) = \\ \nonumber & =
& 4\pi \epsilon R \Big(\frac{\beta \alpha}{\beta - \alpha}\Big)
\frac{L_{eg}}{L_{ee}}\Big(\nabla \mu \Big)R\big(\beta - \alpha\big)=\\
\nonumber & = & 4\pi \beta \alpha
\frac{L_{eg}\epsilon}{L_{ee}}\Big(\nabla \mu \Big) R^2 = C_g C_m
\Big(\nabla \mu \Big) R^2
\end{eqnarray}

where \(C_g=4\pi \beta \alpha\) is a value that only depends on
the geometry of the system and is independent of the values of the
material coefficients, whereas, in contrast,\(C_m=
L_{eg}\epsilon/L_{ee}\) is related only with the values of the
conductivities of the material contained between the plates.

If we distribute the charge \(Q\) uniformly all over the plates we
find the values of the respective charge density \(\sigma_a\) and
\(\sigma_b\) of the inner and outer spheres
\begin{eqnarray}
\sigma_a = \frac{C_g C_m \nabla \mu R^2}{4\pi \alpha^2 R^2}=\frac{C_g C_m \nabla \mu}{4\pi \alpha^2},\\
 \sigma_b
= -\frac{C_g C_m \nabla \mu R^2}{4\pi \beta^2 R^2}=\frac{C_g C_m
\nabla \mu}{4\pi \beta^2}.
\end{eqnarray}

These densities will rotate with the planet with a frequency
\(\nu=(1/\tau)\) where \(\tau\) is the corresponding rotation
period. These charges, rotating with this angular velocity will
generate a dipolar magnetic field. In the case of the internal
plate this magnetic dipole moment is
\begin{eqnarray}\label{campint1}
\nonumber d_a = \int \mathbf{S} d\mathbf{i_a}=\int \mathbf{S}
\sigma_a \nu d\mathbf{A_a}=\\ \nonumber =\int_{0}^{\pi} \pi (R
\alpha \sin{\theta})^2\sigma_a \nu (2 \pi \alpha R \sin{\theta}) R d\theta=\\
= \frac{2 \pi \alpha^2 C_g C_m}{3} \frac{R^4}{\tau}\nabla \mu_k
\end{eqnarray}
while in the case of the outer plate we obtain
\begin{eqnarray}\label{campint2}
d_b =  -\frac{2 \pi \beta^2 C_g C_m}{3} \frac{R^4}{\tau}\nabla
\mu.
\end{eqnarray}

The total magnetic moment will be the sum of (\ref{campint1}) and
(\ref{campint2}) values. Thus the final dipole moment of the
planet is
\begin{eqnarray}\label{magfi1}
d_t & = & d_b +d_a= -\frac{2 \pi (\beta^2-\alpha^2) C_g C_m}{3}
\frac{\nabla \mu R^4}{\tau}.
\end{eqnarray}

The only remaining thing is to return to the value of the gradient
of the chemical potential in terms of the mass and radius of the
planet, given by (\ref{magneticfield}), i.e.
\begin{eqnarray}
\nabla \mu_k = v_k\Bigg(\frac{9GM^2}{8\pi}\frac{1}{R^5}\Bigg).
\end{eqnarray}

If we substitute this value in (\ref{magfi1}) we obtain finally
\begin{eqnarray}\label{campmagneticfinal}
\mathbf{d}_t =  C'_g C'_m G \frac{M^2}{R\tau},
\end{eqnarray}
where we have redefined the geometrical and material factors as
\(C'_g=-3 \pi (\beta^2-\alpha^2) \beta \alpha\), and \(C'_m=
\frac{L_{eg}v_k\epsilon}{L_{ee}}\).

Equation (\ref{campmagneticfinal}) gives the total magnetic field
generated by an object of mass M, radius R and that rotates with a
period \(\tau\).

The resulting magnetic field is a linear combination of two
dipoles: the one produced by the positive charge in the inner
shell and the one produced by the negative charge in the outer
shell. Both shells rotate in the same direction and, as a
consequence, the magnetic fields generated by them are in opposite
directions as they have opposed electric charges. The resulting
dipole is in the direction of the bigger one, in this case in the
direction of that generated by the negative charges because their
linear velocity is higher than that of the positive charges.

At this point we must solve the possibility of the model to
predict a change in the orientation of the generated magnetic
field in order to fit to the experimental data recorded in
basaltic rocks.

For this purpose we have to come back to the thermodynamic
relationship and observe the behavior of the fluxes in the
toroidal component. In this case we have to take on consideration
that forces due to the magnetic field appear in the formulas.
Expressed in tensorial form, constitutive equations are

\begin{eqnarray}
\mathbf{E}=\mathbf{i}/\sigma + \alpha \mathbf{\nabla}T+
R\mathbf{H}\times \mathbf{i}+ N\mathbf{H}\times \mathbf{\nabla}T,
\end{eqnarray}
where \(\mathbf{E}\) and  \(\mathbf{H}\) are the electric and
magnetic field, \(\mathbf{i}\) is the electric current density,
\(T\) is temperature and \(\alpha, \sigma, R, N\) are respectively
the thermal and electrical conductivity and Hall and Nernst
conductivity. In this section we are only interested in the
behavior of this relation in the toroidal component. We could make
some simplifications. Since the system has radial symmetry, the
gradient of temperature and the electric field in this direction
are 0, (\(E_\theta=0, (\nabla T)_\theta=0\)). In the previous
section we considered that the electric current in the radial
direction is 0,(\(j_r=0\)). The magnetic field is in the poloidal
component, (\(\mathbf{H}_r=H_\phi\)). Making these assumptions in
the previous relation we obtain

\begin{eqnarray}
0=j_\theta/\sigma + N H_\phi \nabla T,
\end{eqnarray}
this relation tells us that in the presence of the magnetic field
coming from the radial separation of charges in the core, a
current appears in the azimutal direction due to the gradient of
temperatures. This current is in the opposite direction to that of
the rotation of the planet. As a consequence, the outer shell
"frena" respect to the motion of the planet and this makes the
global magnetic field change in magnitude. If this current is high
enough, the orientation of the magnetic field could change.

\begin{figure}
\label{plotfour}
\includegraphics[scale=0.8]{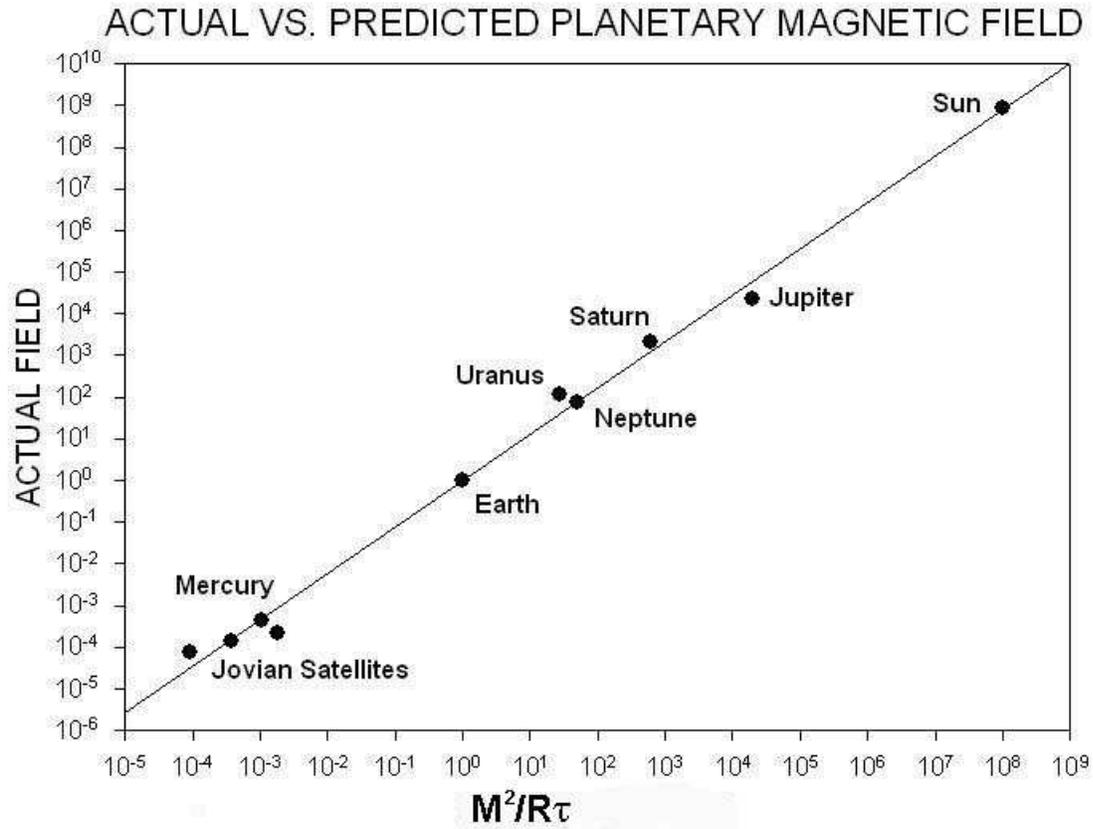}
\caption{Graph for actual vs. predicted magnetic fields for the
planets and satellites with detected global magnetic field. The
actual and predicted field for the Earth is 1 to show that the
scaling predicted by the model agrees with experimental scaling.}
\end{figure}

The process of magnetic field generation by a pressure gradient is
not new. In 1955, \cite{INGLIS} consider this process and
calculated the order of magnitude of the magnetic field. He stated
that the generated field was eighteen orders of magnitude lower
than the actual magnetic field. In the same direction,
\cite{MERRILL1} indicate that this conclusion was predictable
since the pressure gradient effect would not exhibit inversions.
But there are some objections about these conclusions, and we try
to expose them in this section.

The first argument to be objected is that this effect could not
exhibit polarity reversals. In the previous section we propose a
mechanism with which the polarity could be achieved. The mechanism
is the equilibrium between the loss of energy by Joule heating and
the gain in energy due to the orientation of the outer shell
respect to the inner shell.

The second argument needs a more extensive explanation. Inglis, in
his considerations, used statistical mechanics to estimate the
voltage difference between the external and the internal parts of
the Earth's core. He obtained that the pressure in the inner core
would raise the Fermi energy of the iron in \(2 eV\) in relation
to the outer part of the core. The effect in the electrons would
be that these ones tend to accumulate in the outer core because
their energy is lower. The change in the charge density creates an
electric field that pulls the electrons back to the inner core
until equilibrium is reached. At this point, Inglis considered
that an electron needs a voltage difference of \(2 V\) between the
inner and outer part of the core. Although this would be correct
in vacuum, he doesn't considered the fact that these electrons are
in a material medium. As is well known from Sommerfeld theory of
metals, the Fermi energy increase in a metallic wire subjected to
an electric field is (\cite{ASHCROFT, KITTEL}),
\begin{eqnarray}
\delta k = \frac{q}{\hbar}E \tau
\end{eqnarray}
where \(k\) is the Fermi momentum, \(q\) the charge of the
particle (electrons in our model), \(E\) the modulus of the
applied electric field, and \(\tau\) the time between collisions
of the particle in the medium. The time between collisions is
related to the mean free path through the Fermi velocity \(\tau =
\lambda/v_F\), where \(v_F\) is the Fermi velocity. If we use
these relations in the opposite direction, we obtain the value of
the field generated by a shift in the Fermi energy,

\begin{eqnarray}\label{CampFermi}
E=\frac{2 \Delta E_F}{q\lambda}.
\end{eqnarray}

The final result is that the applied field has to be high enough
to give \(\Delta E_F\) of energy in a distance of the order of the
mean free path and not on the complete core thickness. The
difference between both sizes is 14 to 16 orders of magnitude.
Considering that in his discussion he obtained that the field
created was 18 orders of magnitude lower than the actual one, we
can view that this effect could give a considerable field. In the
next section we refine these calculations and obtain the correct
value for the magnetic field.

\cite{STEVENSONPM} also stated that the Ohmic dissipation would be
too large to mantain this effect. In fact there are no currents in
the process of charge separation and so there is not dissipation.
The thermal velocity distribution is isotropic so the amount of
electrons that go inward is the same as the amount of electrons
pointing outward. The electrons going into the core are
accelerated and when they impact with an ion they give the excess
of energy to it by thermalization; in contrast, the electrons
going outward, are deaccelerated and when they impact with an ion
they have less energy than the medium and by the thermalization
process they gain energy. The resulting effect is the absence of
an energy release. The only process that generate heat dissipation
is the one that makes the outer shell tilt respect to the inner
shell, but in this process the velocities are small and the
dissipation is low. In the numerical results section we try to
make an approach to compute this value.

\section{Numerical Results}

Using a similar statistical mechanic approach like the one used by
Inglis we compute the predicted magnetic field for the Earth. We
start by calculating the mean energy for the electrons of the core
under a pressure \(p\). For the pressure we use the value obtained
from hydrostatic equilibrium (\ref{hydrostatic})
\begin{eqnarray}\label{FermiEnergy1}
E_F=\frac{p}{n}=\frac{3 G M^2}{8 \pi R^4}\frac{1}{n},
\end{eqnarray}
were \(n\) is the numerical free electron density of iron in the
core. We use this relation in the Sommerfeld relation obtained in
(\ref{CampFermi}) and substitute the result in the relation
(\ref{CARREGA}) to obtain the amount of charge separated in the
core.

\begin{eqnarray}\label{Q1}
Q=C\Delta R E=4\pi \alpha \beta \epsilon \frac{3 G M^2}{8 \pi
R^2}\frac{\Delta n}{n^2}\frac{1}{q\lambda},
\end{eqnarray}
where q is the fundamental charge and  \(\lambda\) the mean free
path for electrons in the core. Instead of the value of the mean
free path, we use the more usual value of the conductivity. With
this objective we use the Drude-Sommerfeld relation between
conductivity and mean free path, to write
\begin{eqnarray}\label{lambda1}
\lambda = \frac{\sigma m_e v_F}{n
q^2}=\frac{\sigma}{2nq^2}\sqrt{\frac{3Gm_e}{\pi n}}\frac{M}{R^2},
\end{eqnarray}
where \(\sigma\) is the conductivity, \(v_F\) the Fermi velocity,
and  \(m_e\) the electron mass. By substitution of (\ref{Q1}) and
(\ref{lambda1}) on the magnetic dipole moment we obtain,
\begin{eqnarray}\label{CampF}
d =
\frac{2\pi}{3}\frac{R^2}{\tau}Q=\Bigg(q\epsilon\sqrt{\frac{4\pi^3
G}{3m_e}}\Bigg) \alpha \beta \frac{\Delta n}{
n}\frac{\sqrt{n}}{\sigma}\frac{MR^2}{\tau}
\end{eqnarray}

The values we use in the last relation for the conductivity and
the permittivity are,\(sigma = 10^5 \ S/m\) and
\(\epsilon=\epsilon_0 = 8,85\cdot 10^{-12} \ F/m\).

For the value of the increment in the numerical density we assume
that the difference in the numerical density of electrons in the
core is proportional to the difference in the numerical density of
iron atoms so is proportional to the difference in the mass
density
\begin{eqnarray}
\frac{\Delta n}{n} =\frac{\Delta \rho}{\rho}=\frac{(12,8 - 9,9)
 \ g/cm^3}{12,8 \ g/cm^3}=0,22.
\end{eqnarray}

The free electron density at the center of the Earth core will be
\begin{eqnarray}
n_i =\frac{\rho_i}{\rho_o}n_o=\frac{12,8 / g/cm^3}{7,85 \
g/cm^3}\cdot 1,7\cdot 10^{29}=2,77 \cdot 10^{29} \ m^{-3}
\end{eqnarray}
where \(n_i\) and \(n_o\) are respectively the free electron
density in the core and at the surface of the Earth and \(\rho_i\)
and \(\rho_o\) the corresponding mass densities. Using these
values and the values for the mass, radius and period of rotation
of the Earth we obtain the predicted value for the dipole moment
of the Earth,
\begin{eqnarray}\label{CampF}
d = 3,03 \cdot10^{22} A \ m^2.
\end{eqnarray}

 The actual value of the Earth dipole moment is \(7 \cdot 10^{22}
Am^2\). We can see that the value predicted by the model is of the
same order of magnitude as the actual value. This is in
contradiction with the previous results that stated that this
mechanism couldn't give a value close to the experiment.

A remarkable thing that must be noted is that the model obtains
directly the wide known relation about magnetic dipole moment and
the moment of inertia of the object \(d \propto I=MR^2/\tau\).

Returning to the subject of ohmic dissipation we could make an
approach to the value of the power dissipated by a shell moving
respect to the Earth. The less favorable case would be the one in
which the outer shell rotates in opposite sense of the Earth
rotation. In this case the heat generated by ohmic dissipation
would be
\begin{eqnarray}\label{pomega}
P_{\Omega}=\frac{\mathbf{j}}{\sigma}V=\frac{3Q}{(\beta R)^2\tau}.
\end{eqnarray}
where \(\mathbf{j}\) is the current density. In the last relation
we assumed a constant distribution of a charge \(Q\) over a sphere
of radius \(\beta R\) spinning at an angular velocity of \(4 \pi
R/\tau\) respect to the Earth. Substituting \ref{Q1} and
\ref{lambda1} into \ref{pomega} we obtain
\begin{eqnarray}\label{pomega}
P_{\Omega}=144\pi^2 \alpha^2 \beta
\frac{G}{m_e}\frac{\epsilon^2q^2\Delta
n}{\sigma^3}\frac{M^2}{R\tau^2}.
\end{eqnarray}

In the case of the Earth, this value is \(129 \ MW\). This value
is far lower than the values of actual heat dissipation of the
Earth that are of the order of \(TW\).

For the extension of this result to the rest of the planets we
need the values of the conductivity and the free electron density
for all of them. Even though there has been progresses in
calculating the thermodynamical properties of the Earth core and
mantle (\cite{STEVE,LIDUNKA, ALFE1,ALFE2, SHAN, XU}), we have not
yet the detailed values of the transport coefficients of the
objects of the Solar System. In order to obtain such values for
the Solar System, we must make some assumptions about the value of
\begin{eqnarray}\label{Transport}
\frac{\Delta n}{\sigma \sqrt{n}} \simeq
\frac{\sqrt{n}}{\sigma}\simeq \sqrt{n}.
\end{eqnarray}
were we assumed that in the Drude-Sommerfeld model for metals, the
conductivity is directly proportional to the numerical density.
This value is not very different in different planets.
\begin{table}\label{tabplanet1} \centering
\begin{tabular}{|l|l|l|l|l|}
\hline
\textbf{Object} & \(R/R_E\) & \(M/M_E\) & \(\tau/\tau_E\) & \(d/d_E\) \\
\hline
Sun & \(110\) & \(3,33\cdot10^{5}\) & \(28\) & \(10^{7}-10^{8}\)\\
Mercury & \(0.383\) & \(0.0553\) & \(58.8\) & \(3.85 \cdot 10^{-4}\)\\
Venus & \(0.949\) & \(0.815\) & \(243.7\) & \(0 ?\)\\
Earth & \(1\) & \(1\) & \(1\) & \(1\)\\
Moon & \(0.273\) & \(0.0123\) & \(27.4\) & \(0 ?\)\\
Mars & \(0.533\) & \(0.107\) & \(1.03\) & \(0 ?\)\\
Jupiter & \(11.21\) & \(317.7\) & \(0.415\) & \(2\cdot 10^4\)\\
Io  & \(0.286\) & \(0.0150\) & \(1.77\) & \(1.03\cdot 10^{-3}\)\\
Europa & \(0.245\) & \(0.0080\) & \(3.56\) & \(8.97\cdot 10^{-5}\)\\
Ganymede & \(0.413\) & \(0.0248\) & \(7.17\) & \(1.79\cdot 10^{-3}\)\\
Callisto& \(0.378\) & \(0.0180\) & \(16.73\) & \(?\)\\
Saturn & \(9.45\) & \(95.2\) & \(0.445\) & \(605\)\\
Titan& \(0.403\) & \(0.0225\) & \(15.99\) & \(?\)\\
Uranus & \(4.01\) & \(14.5\) & \(0.720\) & \(49.1\)\\
Neptune & \(3.88\) & \(17.1\) & \(0.673\) & \(27.7\)\\
Trito & \(0.212\) & \(0.0036\) & \(5.89\) & \(?\)\\
Pluto & \(0.187\) & \(0.0020\) & \(6.40\) & \(?\)\\
Caronte & \(0.0465\) & \(0.00027\) & \(6.40\) & \(?\)\\
\hline
\end{tabular}
\caption{Table with the values of the radii, mass, rotation period
and magnetic dipole moment of some celestial objects relative to
the Earth.}
\end{table}

From these values we can calculate the ratio (\(d/d_e\)) of the
predicted magnetic dipole moment for the planet \(d\) with the
predicted dipole moment for the Earth \(d_e\), that is, how much
stronger is the field predicted for the object than that predicted
for the Earth.

\begin{table}\label{tabplanet2}
\centering
\begin{tabular}{|l|l|l|l|}
\hline
\textbf{Object} & \(d_p/d_{pE}\)& \(d_r/d_{rE}\) & \(A\)  \\
\hline
Sun & \(3.60\cdot 10^{7}\) & \(10^{7}-10^{8}\)& \(3.6 - 0.36\) \\
Mercury & \(1.35\cdot 10^{-4}\) & \(4 \cdot 10^{-4}\)& \(0.34\) \\
Venus & \(3.01\cdot 10^{-3}\) & \(0 ?\)& \(?\) \\
Earth & \(1\)  & \(1\)& \(1\) \\
Moon & \(1.98\cdot 10^{-5}\)  & \(0 ?\)& \(?\) \\
Mars & \(2.97\cdot 10^{-2}\) & \(0 ?\)& \(?\) \\
Jupiter & \(2.42\cdot 10^{5}\) & \(2\cdot 10^4\)& \(1.21\) \\
Io  & \(4.41\cdot 10^{-4}\)  & \(1.03\cdot 10^{-3}\)& \(0.43\)\\
Europa & \(7.41\cdot 10^{-5}\)  & \(8.97\cdot 10^{-5}\)& \(0.83\)\\
Ganymede & \(2.08\cdot 10^{-4}\)  & \(1.79\cdot 10^{-3}\)& \(0.12\)\\
Callisto & \(5.13\cdot 10^{-5}\)  & \(?\)& \(?\)\\
Saturn & \(2281\)  & \(600\)& \(3.80\) \\
Titan & \(7.86\cdot 10^{-5}\)  & \(?\)& \(?\)\\
Uranus & \(113\) & \(50\)& \(2.26\) \\
Neptune & \(115\) & \(25\)& \(4.6\) \\
Trito & \(1.03\cdot 10^{-5}\)  & \(?\)& \(?\)\\
Pluto & \(3.65\cdot 10^{-6}\)  & \(?\)& \(?\)\\
Pluto & \(2.47\cdot 10^{-7}\)  & \(?\)& \(?\)\\
\hline
\end{tabular}
\caption{Table that shows the predicted value of the magnetic
dipole moment respect the one predicted for the Earth, the
actually observed dipole, and the ratio of both columns, that
shows the correlation between the predicted and the real magnetic
fields. The values for the Earth are 1 to show the agreement of
the actual scaling with the predicted one.}
\end{table}

In table \ref{tabplanet1} are listed the values for the radii,
masses and revolution periods (\(R, M\) and \(\tau\)) of different
objects of the Solar System, relative to the values of the Earth
(\(R_e, M_ e\) and \(\tau_e\)).

The values presented in table (\ref{tabplanet2}) show that the
theory predicts significative intense fields to that objects that
in fact have a significative measured field, while those objects
that have not strong observed fields have a small predicted value.
Moreover, the theory predicts a field strength for the Sun. The
model also predicts global magnetic fields for objects like Venus
and the Moon. If we calculate the magnetic field produced by these
dipoles over the surface we obtain magnetic fields of \(100 \ nT\)
in the surface of Venus and \(19 \ nT\) over the Moon (see table
\ref{tabplanet3}). This values are compatible with the
experimental data from Pioneer Venus Orbiter and Apollo
expeditions. In the case of Venus, the effect of the magnetic
dipole moment could be in superposition with the magnetic effects
coming from ionosphere since the observations of magnetic fields
of several tens of \(nT\)  made by the probe are at the same
altitude as the ionosphere. As a final point we can view that the
model predicts a magnetic field for the great satellites of the
Solar System like the galilean satellites of Jupiter. The fields
predicted also agree with the observed magnetic field
(\cite{OLSON}). Similar magnetic fields are obtained for other
satellites like Titan. Cassini-Huygens mission to Saturn will
measure the magnetic field of Saturn and Titan for which we lack
experimental values from Voyager missions. We also see in figure
\ref{plotfour} the great correlation between the predicted and the
actual value of the magnetic field. In the plot (\ref{plotfour}),
the value for the predicted and existing field for the Earth is
set to 1 to show the agreement between the scaling in the model
and the experimental one.

A final indication must be made about the multipolarity and the
dynamic of the model. A spherical shell has higher order
multipoles that are not computed here because it is out of the
scope of an introductory paper, numerical calculations with the
multipole distribution will be obtained shortly by the aid of
computer models. Once multipolarity will be obtained, dynamics
will be added by the introduction of magnetic interaction between
both shells and with an outer field (planetary field in the case
of a satellites or solar field in the case of a planet) and
variability of pressure and temperature in the core of the object.

\begin{table}\label{tabplanet3}
\centering
\begin{tabular}{|l|l|}
\hline
\textbf{Object} & \(B_s \ (nT)\)\\
\hline
Sun & \(923000\)\\
Mercury & \(72\)\\
Venus & \(101\)\\
Earth & \(30000\)\\
Moon & \(30\)\\
Mars & \(4200\)\\
Jupiter & \(465000\)\\
Io  & \(9100\)\\
Europa & \(2400\)\\
Ganymede & \(1400\)\\
Callisto & \(460\)\\
Saturn & \(76000\)\\
Titan & \(570\)\\
Uranus & \(34000\)\\
Neptune & \(57000\)\\
Trito & \(520\)\\
Pluto & \(17\)\\
Caronte & \(73\)\\
\hline
\end{tabular}
\caption{Table that shows the predicted value of the magnetic
field over the surface of the planet. This value is calculated
from table (\ref{tabplanet2}) with (\(d_r=A d_{rE}\)) and using a
value of \(d_{rE}=7\cdot 10^{22} Am^2\) for the Earth}
\end{table}

\section{Conclusions}

The model presented in this paper describes, in a single theory,
the magnetic field detected in several objects of the Solar
System. The values obtained are close to the experimental ones.
The model proposed could be used to break the initial symmetry of
the system in the numerical geodynamo models. These models
predict, also, a tilt of the magnetic dipole moment respect to the
axis of rotation. This tilt could depend on the properties of the
materials that fill the shell, like conductivity. Moreover, we can
see that following this proposal, the planets with higher tilts of
the magnetic dipole have higher quadrupole moments, because the
total field would be generated by two crossed dipole magnets. This
feature of the field agrees with the observed magnetic field of
Uranus and Neptune (\cite{CONNEREY1, CONNEREY2}) and with the fact
that the magnetic field of Saturn has a lower quadrupole moment
(\cite{CONNEREY3}). The model can also explain the observation
that the quadrupolar component in the Earth magnetic field becomes
more important in the transition between normal and reversed
polarity of the field (\cite{SCHNEIDER}). As the field is
generated by the effect of two concentric spherical shells,
higher-order multipoles are predicted by the theory (a rotating
sphere is not only dipolar or quadrupolar). The calculation of
these higher orders is left because it is out of the scope of an
introductory paper. The only planet unexplained by this model is
Mars, but notice that the zero field could be an extreme case
where the outer and the inner shell have opposite magnetic dipole
values.

Magnetic fields for the galilean satellites are also predicted.
Some theories say that the fields for these satellites are
explained by the Jupiter magnetic field induction, and the reason
for this hypothesis is that the fields of the satellites change of
direction following the changes of Jupiter's field. This reason is
not enough to think that the field is induced. The field could be
created as is explained in the model and, once created, it moves
like a compass near a magnet.

The article also calculates the composition dependence of the
model. The result is that the field depends of the inverse of the
root of the numerical density. This dependence shows that the
influence of the composition is not too high because increasing 10
times the numerical density only generates a multiplication of 0.3
of the magnetic field. Even though the composition of the nuclei
of the several planets could be very different, its numerical
densities couldn't variate enough to change the order of magnitude
of the generated fields.

The model could be generalized to other objects in the Universe
like the main sequence stars and the neutron stars, objects where
the gravitational effects are higher than in the planets. Even
though these objects are surely not metallic in its core, charged
particles could be found like electrons that could behave like the
electrons in the model. In objects like neutron stars, geodynamo
is surely less probable due to the ultra high density of the
entire object.

In conclusion, the theory achieves with a thermodynamic model a
scaling law that agrees with the values of the Solar System
magnetic fields. Even though the accepted theory for the
generation of the magnetic fields is the geodynamo, this
thermodynamic model could be used to explain the order of
magnitude and the scaling of the actual fields, fact that couldn't
be explained with the current theory.


\end{document}